\documentstyle[preprint,eqsecnum,aps,floats]{revtex}
\begin{document}
\input epsf
%%%%%%%%%%%%%%%% READ TH%%%%%%%%%%%%%%%% READ THIS %%%%%%%%%%%%%%
%%%%%%%%%%%%%%%%%%%%%%%%%%%%%%%%%%%%%%%%%%%%%%%%%%%%%%%%%%%%%%%%%%%%%
%   This paper uses REVTEX Version 3.0
%
%%%%%%%%%%%%%%%%%%%%%%%%%%%%%%%%%%%%%%%%%%%%%%%%%%%%%%%%%%%%%%%%%%%%%%%%%%%
\begin{titlepage}
\preprint{NSF-ITP-96-141, UCSBTH-96-26}
\title{Quantum Cosmology:\\
Problems for the 21$^{\rm st}$ Century\thanks{To appear in {\sl Physics
2001}, ed.~by M.~Kumar and
in the {\sl Proceedings of the 10$^{\rm th}$ Yukawa-Nishinomiya
Symposium}, November 7--8, 1996, Nishinomiya, Japan.}}
\author{James B.~Hartle\thanks{ Institute for
Theoretical Physics, University of California, Santa Barbara,
CA, 93108; e-mail address:
hartle@itp.ucsb.edu}}

\date{\today}
\maketitle

\tighten
\begin{abstract}

Two fundamental laws are needed for prediction in the universe: (1) a
basic dynamical law and (2) a law for the cosmological
initial condition. 
Quantum cosmology is the area of basic research concerned with
the search for a theory of the initial cosmological state. The issues
involved in this search are presented in the form of eight problems.

\end{abstract}

\end{titlepage}

\tighten

\section{What are the Fundamental Laws?}

Physics, like other sciences, is concerned with explaining and
predicting the regularities of specific physical systems. Stars, the
solar system, high-temperature superconductors, fluid flows, atoms, and
nuclei are just some of the many examples. Beyond particular
systems, however, physics aims at finding laws that predict the regularities
exhibited universally by {\it all} physical systems --- without exception, without
qualification, and without approximation. These are the {\it
fundamental} laws of physics. This essay is concerned with the
fundamental law for the initial condition of the universe.

Ideas for the nature of the fundamental laws have 
varied as new realms of phenomena have been explored experimentally.  
However, until recently, all of the various ideas for fundamental laws have had
one feature in common: They were proposals for dynamical laws --- laws
that predicted regularities in time.  The laws of Newtonian mechanics,
electrodynamics, general relativity, and quantum theory all have this
character.

The Schr\"odinger equation is an example of fundamental dynamical law:
\begin{equation}
i\hbar\ \frac{\partial \Psi}{\partial t} = H\Psi\ .
\label{oneone}
\end{equation} 
A fundamental theory of dynamics 
supplies the Hilbert space and the Hamiltonian
operator $H$. However, a differential equation like (\ref{oneone}) makes
no predictions by itself. 
To solve (\ref{oneone}), an initial condition ---  the state
vector at one moment --- must also be given. 
The Schr\"odinger equation then expresses the regularities in time
that emerge from this initial state.

Where do the boundary conditions necessary to solve dynamical laws come
from? In most of physics we study the evolution of subsystems of the
universe and
determine the boundary conditions by observation or experimental
preparation. If we are interested in the evolution of the
electromagnetic field in a room and observe no incoming radiation, we
solve Maxwell's equations with no incoming radiation boundary
conditions. To predict the probability for the decay of an atom prepared
in an excited state, we solve the Schr\"odinger equation with that
excited state as an initial condition at the time of preparation, and so on. Boundary conditions
for the evolution of subsystems are obtained from {\it observations}
 of the rest
of the universe outside the subsystem of interest.

Cosmology, however, presents us with  an essentially different problem. 
The dynamical laws governing the evolution of the
universe --- the classical Einstein equation, for instance --- require
boundary conditions to yield solutions. But in cosmology, by
definition, there is no ``rest of the universe'' to pass their
specification off to.  The cosmological boundary condition must be one of the
fundamental laws of physics.

The inference is inescapable from the physics of the last sixty years that
the fundamental laws are all quantum mechanical. If that is assumed,  
a theory of the initial
condition is a theory of the universe's
initial quantum state. The
search for a fundamental theory of this initial cosmological quantum
state is the aim of that area of basic research which has come to be
called quantum cosmology.

\eject
A view thus emerges that there are two fundamental laws of physics:
\begin{itemize}
\item A theory of the basic dynamics,
\item A theory of the initial condition of the universe.
\end{itemize}
Were the universe governed by the Schr\"odinger equation (\ref{oneone}),
the basic theory of dynamics would specify the Hamiltonian
$H$; a theory of the initial condition would be a law for the initial
quantum state.

The search for the fundamental dynamical law has been seriously under
way since the time of Newton. Classical mechanics, Newtonian gravity,
electrodynamics, special relativity, general relativity, quantum
mechanics, quantum electrodynamics, quantum chromodynamics, 
electro-weak theory, grand unified theories, and superstring
theories are but some of the important milestones in this search. By
contrast, the search for a theory of the initial condition of the
universe has been
seriously under way for only a few decades. Why this difference? The
answer lies in the empirical locality of the fundamental interactions on scales
above the Planck length ($\sim 10^{-33}$cm), or put differently, the empirical fact 
that the fundamental interactions may be effectively described
by a local quantum field theory on these scales. Assuming locality, the
Hamiltonian of the whole universe can be deduced from experiments on
familiar, laboratory, scales.
However, typical ideas for the initial
quantum state of the universe are non-local.  They imply regularities {\it in space}
that emerge mostly on large, cosmological scales. For example, the
temperature of the cosmic microwave background is the same across the
sky to one part in $10^5$, a distance which corresponded to $10^{20}$km
at the time the radiation was emitted.  It is only the recent
progress in observational cosmology that has given us a picture of the
universe on large enough scales of both space and time that is sufficiently
detailed to confront with the predictions of a theory of the initial state
of the universe.

\section{Quantum Cosmology and The Everyday}

Can those not interested in regularities on cosmological scales safely
ignore the initial condition of the universe? Not if they seek a
fundamental explanation of a number of its features we ordinarily take for
granted. In this section we offer a few examples.

\noindent{\bf $\bullet$ Isolated Subsystems}

In one way we use a very weak theory of the initial condition every day.
Many subsystems of the universe, in the laboratory and elsewhere, 
 are approximately isolated for periods of
time and can can be approximately described by solving the
Schr\"odinger equation for the subsystem alone.  In effect, we assume
that for the purposes of making predictions about the subsystem, the wave
function of the universe can be approximated by
\begin{equation}
\Psi (q^i, Q^A, t) \approx \psi (q^i, t) 
\Phi (Q^A, t)
\label{twoone}
\end{equation}
where $q^i$ and $Q^A$ are co\"ordinates referring to subsystem and the
rest of the universe respectively and $\psi$ and $\Phi$ evolve {\it
separately} under the Schr\"odinger equation.  But what are the grounds
for such an approximation? They do not lie in the nature of the
Hamiltonian because that generally specifies interactions between all
the co\"ordinates. Rather the existence of isolated
subsystems is a property of the quantum state. In discussing isolated
subsystems, we are making weak quantum
cosmological assumptions about the nature of this initial state.

\noindent{\bf $\bullet$ The Quasiclassical Realm}

Classical deterministic laws govern a wide range of phenomena in the
universe over a broad span of time, place, and scale. This
quasiclassical realm is one of the most immediate facts of our
experience.  But indeterminacy and distributed probabilities are the
characteristics of a quantum mechanical universe. Classical deterministic
dynamics 
can be but an approximation to the unitary
evolution of the Schr\"odinger equation and the reduction of the state
vector.  To what do we owe the validity of this approximation? In part
it arises from a coarse-grained description with positions and
momenta specified to accuracies well above the limitations of the
uncertainty principle for instance. But coarse graining is not enough; there
must also be some restriction on the initial state.  Ehrenfest's theorem is a
simple illustration of why. For the motion of a particle in
one dimension, Ehrenfest's theorem
relates the acceleration of the expected position to
the expected value of the force:
\begin{equation}
m\ \frac{d^2\langle x\rangle}{dt^2} = - \left\langle\frac{\partial
V(x)}{\partial x}\right\rangle \ .
\label{twotwo}
\end{equation}
This is generally true, but for {\it certain} states, typically narrow
wave packets, the right hand side may be replaced by the force evaluated
at the expected position to a good approximation resulting in the
deterministic classical equation of motion
\begin{equation}
m\ \frac{d^2\langle x\rangle}{dt^2} \approx -
\frac{\partial
V(\langle x \rangle)}{\partial x} \ .
\label{twothree}
\end{equation}
Just as only certain states lead to classical behavior in this simple
model, so also only certain cosmological
initial conditions will lead to the quasiclassical realm of familiar
experience.\footnote{For a more quantitative discussion see,
{\it e.g.}~\cite{Har94b}.} That too is a feature of the universe that must
ultimately be traced to the initial condition.

\noindent{\bf $\bullet$ Homogeneity of the Thermodynamic Arrow of Time.}

Isolated systems evolve towards equilibrium. That is a consequence of
statistics. But in this universe
presently isolated systems are mostly evolving towards equilibrium in
the {\it same} direction of time. That is the homogeneity of the thermodynamic
arrow of time. This is not a fact which can be explained by statistics
or a property of the Hamiltonian alone for that is approximately
time-reversal invariant. 
The homogeneity of the thermodynamic arrow of time follows from a
fundamental law of the initial condition which mandates that the
progenitors of today's isolated systems were all far from equilibrium in
the early universe.
As Boltzmann put it: ``The second law of
thermodynamics can be proved from the [time-reversible] mechanical
theory if one assumes that the present state of the universe ... started
to evolve from an improbable state'' \cite{Bol97}. 

\noindent{\bf $\bullet$ History.}

The reconstruction of history is useful for understanding the present in
science as well as in human affairs.  For example, we can best understand
the character of biological species by understanding their evolution. 
We can best explain the present large scale distribution of galaxies
by understanding how galaxies arose from tiny density fluctuations
present shortly after the big bang. 
Such examples could be easily multiplied.

In physics, the reconstruction of history means using the fundamental
laws to calculate the probabilities of alternative past events assuming
the values of present records.  Classically, present records {\it alone}
are enough to calculate those probabilities by 
using them as the starting point for running the deterministic classical
equations of motion backward in time.  To reconstruct history in quantum
mechanics, however, requires a theory of the initial condition {\it in
addition} to present records.

The source of this difference between classical and quantum mechanics
can be traced to the {\it arrow of time} in usual quantum
theory.\footnote{There are generalizations of quantum theory without an
arrow of time in which the asymmetry of the usual theory may be
understood as a difference between initial and final conditions,
{\it e.g.}~\cite{GHxx}. We shall not consider these here.} Quantum mechanics
treats the future differently from the past.  To be sure, the
Schr\"odinger equation (\ref{oneone}) can be run backwards in time as
well as forwards.  But the Schr\"odinger equation is not the only law of
evolution in quantum theory.  In the usual story, when a measurement is made,
the wave function is ``reduced'' by the action of the projection operator
$P$ representing the outcome of the measurement, and then renormalized.
This is a ``second law of evolution'':
\begin{equation}
\Psi \rightarrow \frac{P\Psi}{||P\Psi||}\ .
\label{twofour}
\end{equation}
The evolution of the Schr\"odinger equation forwards in time is interrupted by
(\ref{twofour}) on a measurement. While the Schr\"odinger equation can
be run backwards in time, the law (\ref{twofour}) cannot, and that is a
simple way of seeing the arrow of time in usual quantum mechanics. The
same kind of arrow of time persists in more general quantum theories of
closed systems where (\ref{twofour}) is effectively used in in the
construction of the probabilities of histories which are not 
necessarily of the outcomes of measurements.

How then does one calculate the probabilities of past events assuming
present records in quantum mechanics? The simple answer is that one
works forwards in time from the initial state. Evolving forwards using
(\ref{oneone}) and (\ref{twofour}) one calculates the joint probabilities 
of {\it both} alternative events of interest
in the past {\it and} the alternative values of the present records that
follow them. From these one calculates the conditional probabilities of
the past events given our particular present records in the usual way.

This process involves the initial state in an essential way. Strictly
speaking, therefore, one cannot make any statements
about the past without a theory of the universe's initial condition. 

\noindent{\bf $\bullet$ Phenomenology of the Initial Condition}. 

While the above four everyday features of the universe are fundamentally
traceable to the universe's initial quantum state, there is a large set
of initial states that would give rise to them. Put differently,
the existence of isolated subsystems together with the applicability of
classical physics, the second law of thermodynamics, and the possibility
of historical
explanation are not strong constraints on the initial quantum state.
Neither are the observations of large scale features of the universe
such as its approximate homogeneity and isotropy or the fluctuations in
the cosmic background radiation. The data are meager and the Hilbert
space of the observable universe is vast.

It would be possible to investigate quantum cosmology
{\it phenomenologically} by asking for the constraints present observations
place on the initial
state of the universe.  A density matrix  $\rho$ is
the   way quantum mechanics represents the statistical distribution of states
with associated probabilities that would be inferred. 
To investigate the
initial condition phenomenologically is therefore to ask for the density
matrices consistent with observed features of the universe.

The observed features of the universe may not uniquely fix an initial
condition but one should 
not exaggerate their weakness. The  density matrix
$\rho = I/Tr(I)$, where $I$ is the unit matrix, is the unique representation
of complete 
ignorance of the initial condition ({\it i.e.} no condition at all).  But it also corresponds to infinite temperature in
equilibrium $(\rho \propto  \exp (-H/kT))$ --- an initial condition whose
implication of infinite temperature today is obviously inconsistent with
present observations.

The entropy $S/k = -Tr(\rho\log\rho)$ is a measure of the missing
information about the initial state in a density matrix $\rho$.
Most of the entropy in the matter in the visible universe is in the
cosmic background radiation, a number of order $S/k \approx
10^{80}$. As Penrose \cite{Penxx} has stressed, this is a large number,
but infinitesimally small compared to the maximum possible value of
$S/k \approx 10^{120}$ if all that matter composed a black hole.

This essay, however, is not concerned with phenomenology. {\it Rather,
it is concerned with the fundamental law of the
initial condition. We shall therefore assume that the universe has a
initial state $|\Psi\rangle$ and discuss the issues involved in a search
for the principles which determine it.}

\section{Problems}

Enumerating issues is one way of summarizing the present status of an
area of science, and motivating future research. Certainly, setting
problems is a more pleasant task than solving them, and quantum cosmology
is such a young field that it is easier to summarize problems than to survey
accomplishments. It is in this spirit
that the author offers the following eight problems in quantum cosmology:

\noindent {\bf $\bullet$ Problem 1: What Principle Determines the 
Initial Condition of the
Universe?}
 
The evidence of the observations is that the universe was simpler
earlier than it is now --- more homogeneous, more isotropic, with matter
more nearly in thermal equilibrium. This is evidence for a simple,
discoverable initial condition of the universe.  
But what principles determine that initial state?

The most developed proposal for a principle determining the initial
condition is the ``no-boundary'' wave function of Stephen Hawking and
his associates \cite{Haw84}. The idea is that the initial condition of a closed
universe is the cosmological analog of a ground state.  
This does not mean 
the lowest eigenstate of some Hamiltonian.  Intuitively, the
total energy of a closed universe is zero for there is no place outside
from which to measure it. Correspondingly the Hamiltonian
vanishes.

But the lowest eigenstate of a Hamiltonian is not the only way to find
the ground state even in the elementary case of a particle
moving in a potential $V(x)$. In that case, the ground state wave function
may be expressed directly as a
sum over Euclidean paths, $x(\tau)$:
\begin{equation}
\psi_0(y) = \sum_{{\rm paths} \ x(\tau)} \exp\Bigl(-I\bigl[x(\tau)\bigr]/\hbar\Bigr)
\label{threeone}
\end{equation}
where $I=\int d\tau [m\dot x^2/2 + V(x)]$ is the Euclidean action. The 
sum is over paths $x(\tau)$ that have the argument of the wave function, $y$,
as one end point, and a configuration of minimum action in the infinite
past as another. Verify it for the harmonic oscillator for example!

This construction of a ground state wave function generalizes to closed
universes.  For definiteness suppose, for a moment, that the basic
variables of the fundamental dynamical theory are the geometry of
four-dimensional spacetime ${\cal G}$, represented by metrics on manifolds,
together with matter fields such as the quark, lepton, gluon, and Higgs
fields which we generically denote by $\phi(x)$. 
The arguments of cosmological wave
functions are these basic variables restricted to spacelike surfaces,
specifically the three-geometries of these surfaces, $^3{\cal G}$,  and the
field configurations on these surfaces, $\chi({\bf x})$. The ``no-boundary'' wave
function is of the form
\begin{equation}
\Psi_0 [^3{\cal G}, \chi ({\bf x})] = \sum_{{\cal G}, \phi(x) \in C} 
\exp\Bigl(-I\bigl[{\cal G}, \phi (x)\bigr]/\hbar\Bigr)
\label{threetwo}
\end{equation}
where $I[{\cal G}, \phi ({\bf x})]$ is the action for gravitation
and matter. The ``no boundary'' wave function is specified by giving the
class $C$ of four geometries ${\cal G}$, and matter fields $\phi(x)$ 
summed over in (\ref{threetwo}). So that the construction is
analogous to (\ref{threeone}), these geometries ${\cal G}$ should have 
Euclidean (signature $++++$) and have one boundary at which they
match the three-geometry where the wave function is evaluated. 
The matter fields must similarly match their boundary value. 
The defining requirement is
that the ${\cal G}$'s have no {\it other} boundary, whence the name
``no-boundary'' proposal.

Nothing goes on in a typical ground state in a fixed background
spacetime. In field theory, the ground state is
the time-translation invariant vacuum! However, this is not the context
of the quantum cosmology of closed universes. Spacetime
geometry is not fixed and there is
therefore no notion of time-translation. Interesting histories therefore
can happen; and the attractive nature of gravity makes things happen even
in this cosmological analog of the ground state.  In particular, initial,
small, quantum, ground state fluctuations from homogeneity and isotropy
that are predicted by this initial condition can grow by gravitational 
attraction
to produce all the complexity in the universe that we see today.

This prescription for the ``no-boundary'' wave function is not 
complete.  The reason is that the action $I[{\cal G}, \phi (x)]$ for
gravitation coupled to matter is unbounded below. Were the sum in
(\ref{threetwo}) extended over real, Euclidean geometries and fields,
it would diverge! Rather, 
the sum must be taken over a class $C$ of
{\it complex} geometries and fields.  A complex contour of summation
is, in fact, essential
for the ``no-boundary'' wave function to predict the nearly classical
behavior of geometry we observe in the present epoch.  But many
different complex convergent contours are possibly available and
correspondingly there are many different ``no-boundary'' wave functions.
These do not differ in their semi-classical predictions; but we still
lack a complete principle for
fixing this wave function of the universe.

The ``no-boundary'' idea has been described in terms of an effective
theory of dynamics in which spacetime and matter fields are treated as
fundamental variables. If spacetime is not fundamental, as in string theory
or non-perturbative quantum gravity, then extending the idea to such
theories becomes an important question. The essentially topological
nature of the idea gives some hope that such an extension is possible. 

The ``no-boundary'' wave function is not the only idea for a theory of
the initial condition. Other notable candidates are the ``spontaneous
nucleation from nothing wave function'' \cite{Vilsum}, and the ideas
associated with the ``eternally self-reproducing inflationary universe''
\cite{Linsum}.  Space
does not permit a review of these and other theories, and the similarities
and differences in their predictions.  Discriminating between these and
other ideas that may arise is certainly a problem for the 21st century.

\noindent{\bf $\bullet$ Problem 2: How Can Quantum Gravity be Formulated for
Cosmology?}

Gravity governs the evolution of the universe on the largest scales of
space and time. That fact alone is enough to show that a quantum theory
of gravity is required for a quantum theory of cosmology. Were the
behavior of the universe on present cosmological scales all that was
of interest, then a low energy approximation to quantum gravity 
would be
adequate.  Indeed most of the exploration of quantum cosmology has been
carried out in such a low energy approximation assuming spacetime geometry and
quantum fields are the basic variables with Einstein's theory
coupled to matter
as the basic action. Any divergences that arise are truncated 
in one way or another.

It is a reasonable expectation that low-energy, large scale, features of
the universe, such as the galaxy-galaxy correlation function, are
insensitive to the nature of quantum gravity on very small scales. But
in quantum cosmology we aim not only at an explanation of such large
scale features, but also at a theory of the initial condition
adequate to describe the probabilistic details of the earliest moments of the universe. The
inevitability of an initial singularity in classical Einstein
cosmologies strongly suggests that the earliest moments of the universe
will exhibit curvatures of spacetime characterized by the Planck length
\begin{equation}
\ell \equiv (\hbar G/c^3)^{1/2} \approx 10^{-33}{\rm cm}
\label{threethree}
\end{equation}
--- the only combination of the three fundamental constants governing
relativity, quantum mechanics, and gravity that has the dimensions of
length. By making similar combinations with the right dimensions we
can exhibit the Planck scales of energy and time. The universe at the epochs 
characterized by these scales will therefore depend on the
detailed form of the fundamental quantum dynamical law for gravity.

There are a number of candidates for a finite, manageable quantum theory
of gravity, notably superstring theory and non-perturbative canonical
quantum gravity. However, neither of these theories is ready for
application to quantum cosmology. String theory, for instance, exists in
a practical sense as a set of rules for classical backgrounds and quantum
perturbations away from them.
Developing such theories to the point where they can be used for the
non-perturbative  quantum dynamics of closed cosmologies is thus an
important problem.

The problem to be faced is not merely one of technique. Both of the
approaches mentioned, and others as well, hint that spacetime geometry
may not be a basic dynamical variable. 
If that is true, it becomes a conceptual issue just how to
frame cosmological questions in the variables of the fundamental
dynamical theory.
\eject

\noindent{\bf $\bullet$ Problem 3: What is the Generalization of 
Quantum Mechanics
Necessary for Quantum Gravity and Quantum Cosmology?}

A generalization of usual quantum mechanics is needed for quantum gravity.  
That is because usual quantum mechanics relies in essential ways 
on a fixed, background
spacetime geometry,  in particular, 
to specify the notion of time that enters centrally into
the formalism.
This reliance on a fixed notion of time shows up
in any of the various ways of formulating usual quantum theory --- the
idea of a state at a moment of time, the preferred role of time in the
Schr\"odinger equation, the inner product at a moment of time, the
reduction of the state vector at a moment of time, the commutation of
fields at spacelike separated points, the equal time commutators of
co\"ordinates and momenta, {\it etc., etc}.

But in quantum gravity, spacetime geometry is not fixed, rather it is a
quantum dynamical variable, fluctuating and generally without definite
value.  A generalization of usual quantum theory that does not require a
fixed spacetime geometry, but to which the usual theory is a good
approximation in situations when the geometry is
approximately fixed, is therefore needed for quantum gravity and quantum
cosmology. What, therefore, do we mean more 
generally by a quantum mechanical theory?

The most general objective of any quantum theory is the prediction of
the probabilities of alternative, coarse-grained histories of
the universe as a single, closed quantum mechanical system.  For
example, one might be interested in predicting the probabilities of the
set of possible orbits of the earth around the sun. Any orbit is
possible, but a Keplerian ellipse has overwhelming probability. Such
histories are said to be coarse-grained because they do not specify the
co\"ordinates of every particle in the universe, but only those of the
center of mass of the earth and sun, and these only crudely and not at every time.

However, the characteristic feature of a quantum mechanical theory is
that consistent probabilities cannot be assigned to every set of
alternative histories because of quantum mechanical interference.
Nowhere is this more clearly illustrated than in the famous two-slit
experiment shown in Figure 1. Electrons can proceed from an
electron gun at left towards detection at a point $y$ on a screen
along one of two possible histories --- the history passing through the upper
slit, $A$,  and the history passing through the lower slit, $B$. In the usual story,
probabilities cannot be assigned to these two
histories if we have not {\it measured} which slit the electron passed
through.  It would be inconsistent to do so because the 
the probability to arrive at $y$ would not be the sum
of the probabilities to arrive there going through the upper slit and
lower slit:
\begin{equation}
p(y) \not= p_A (y) + p_B (y)\ 
\label{threefour}
\end{equation}
because of quantum mechanical interference. 
In quantum mechanics probabilities are squares of amplitudes 
and, of course,
\begin{equation}
|\psi_A (y) + \psi_B (y)|^2 \not= |\psi_A (y)|^2 + |\psi_B (y)|^2\ .
\label{threefive}
\end{equation}
A necessary consistency condition would not be satisfied. 

\begin{figure}
\centerline{\epsfysize=3.00in \epsfbox{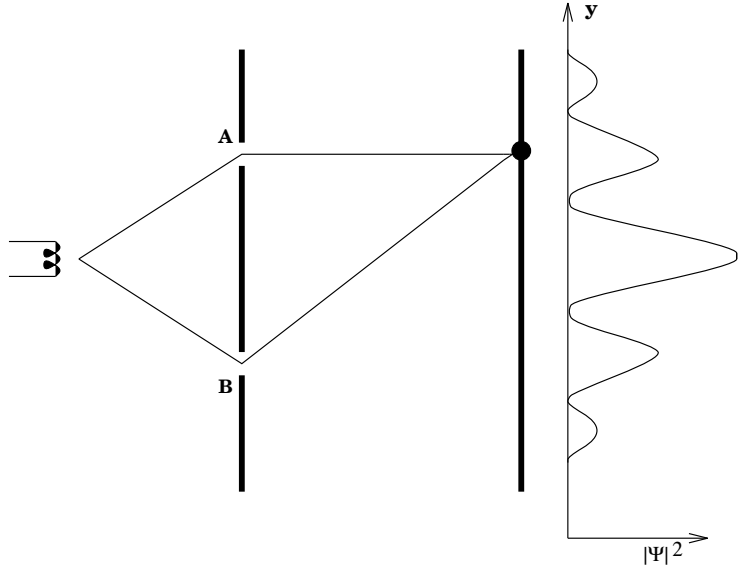}}
\vskip .26 in
\centerline{\caption{\bf The Two-Slit Experiment}}
\label{figone}
\end{figure}

A rule is  thus needed in quantum theory to specify which sets of
alternative histories may be assigned probabilities and which may not.
In the usual, ``Copenhagen'' formulations of quantum mechanics presented
in textbooks, probabilities
can be assigned to the histories of alternatives of a subsystem 
that were ``measured''
by an ``observer''. But such formulations are not general
enough for quantum cosmology which seeks to describe the early
universe where there were neither measurements nor observers present. 

In the more general quantum mechanics of closed systems\footnote{
For expositions see \cite{Har93a,Omn94}.} that rule is simple:
probabilities can be assigned to just those sets of alternative histories for which
there is vanishing interference between the individual members as a
consequence of the system's initial state $|\Psi\rangle$. 
To make this quantitative we need the 
measure of this interference.

When there is a well-defined fixed notion of time, sequences of
alternative sets of events at a series of times define  a set of
alternative histories. 
 An individual history in such a set is a particular series of events, say
$\alpha \equiv (\alpha_1, \alpha_2, \cdots, \alpha_n)$ at times $t_1 < t_2
< \cdots < t_n$. In usual quantum mechanics  such a
history is represented by a corresponding chain of (Heisenberg-picture) operators,
\begin{equation}
C_\alpha \equiv P^n_{\alpha_n} (t_n) \cdots P^2_{\alpha_2} (t_2)
P^1_{\alpha_1} (t_1) \ ,
\label{threesix}
\end{equation}
time ordered from right to left. The application of the $C_\alpha$ to the
initial state vector $|\Psi\rangle$ gives the branch state vector
\begin{equation}
C_\alpha |\Psi\rangle
\label{threeseven}
\end{equation}
corresponding to the history. Interference vanishes in a set of alternative
histories when the branch state vectors corresponding to the different
histories are mutually orthogonal. Sets of 
alternative histories with vanishing interference are said to {\it
decohere}. The probabilities $p(\alpha)$ of the individual histories in
a decoherent set are the squared lengths of the branch state vectors
\begin{equation}
p(\alpha) = || C_\alpha |\Psi\rangle ||^2 \ .
\label{threeeight}
\end{equation}
Decoherence insures the consistency of these probabilities. 

Interference is thus measured by the {\it decoherence
functional}:
\begin{equation}
D(\alpha^\prime, \alpha) = \left\langle\Psi|C^\dagger_{\alpha^\prime}
C_\alpha|\Psi\right\rangle
\label{threenine}
\end{equation}
which becomes the central element in the theory. The condition of
decoherence and the resulting probabilities may be expressed by the
single formula
\begin{equation}
D(\alpha^\prime, \alpha) = \delta_{\alpha^\prime\alpha}  p(\alpha)\ .
\label{threeten}
\end{equation}

The sets of possible coarse-grained histories, their decoherence
functional, and (\ref{threeten}) are the minimal elements of a
quantum theory. A broad framework for quantum theories built on
these elements, called generalized quantum 
mechanics, can be formulated in terms of decoherence functionals obeying
general principles of Hermiticity, normalization, positivity, and the
principle of superposition \cite{Ish94,Har95c}.

Histories represented by strings of projections at definite moments of
time (\ref{threesix}) and a decoherence functional (\ref{threenine}) are the
way that usual quantum theory implements the principles of
generalized quantum theory. But there are many other ways, and among
them are possibilities for generalizing usual quantum mechanics so that
it works in the absence of a fixed spacetime geometry.  Generalized
sum-over-histories quantum theories
have been discussed that put quantum theory into
fully spacetime form with four-dimensional notions of histories, coarse
grainings, and decoherence \cite{Har95c}.  But the principles of  generalized quantum
mechanics are only a minimal set of requirements for quantum theory.
What further principles determine the correct quantum mechanics for
quantum gravity and quantum cosmology?

\noindent{\bf $\bullet$ Problem 4: What are the Definite Predictions of the
Initial Condition for the Universe on Large Scales?}

Extracting the predictions of a theory of the initial condition and
comparing them with observations is a central problem in quantum
cosmology. Predictions take the form of probabilities for present observations. 
The theory stands or falls on those predictions with
probabilities sufficiently close to one (or zero) being observed (or not
observed).  These are called the {\it definite}
predictions of the theory. We expect few of them.  A simple,
comprehensible, discoverable theory of the initial condition cannot
predict all the complexity observed in the present universe with
probability near one\cite{Har96b}.  Rather, most predictions, as of the stock 
market,
the weather, or the number of moons of Jupiter,  will have more 
distributed probabilities 
based on the initial condition alone. (The vast majority will be
nearly uniformly distributed which is no prediction at all.)  In quantum
cosmology one must search among the possible predictions for those which
are predicted with probability near one.  
Interestingly, definite predictions may occur on
all scales.  For the purposes of simplicity we have divided the problem
of what are the definite predictions of a theory of the initial condition
into problems concerning regularities on cosmological, familiar, 
and microscopic scales.

Quantum cosmologists expect that a number of the general large scale
features of the universe will be definite predictions of a theory
of its initial condition. These include an approximately classical
cosmological spacetime geometry after the Planck epoch, the 
approximate homogeneity and isotropy of the geometry and matter
on scales above several hundred megaparsecs\footnote{ The megaparsec
is a convenient unit for cosmology. One megaparsec is 3.3 million 
light years. The size of the visible universe is several thousand
megaparsecs.}, the approximate spatial flatness of the universe (or
what is the same thing its vast age in Planck units), the initial
spectrum of quantum fluctuations which grew to become the galaxies,
a sufficiently long inflationary epoch, and the cosmological abundances
of the matter and radiation species. 

The probabilities for these features of the universe arising from 
various theories of the initial condition have been explored in 
highly simplified models valid only in limited regions of the configuration
space of possible present universes. The output of some of these
calculations, such as the prediction of the spectrum of initial 
quantum fluctuations \cite{HHaw85},  are among the most successful achievements
of quantum cosmology. But much more needs to be done to extend these
calculations to the whole of configuration space 
with greater accuracy, generality and a precise 
quantum mechanical interpretation. That is
a practical and immediate problem for quantum cosmology.

\noindent{\bf $\bullet$ Problem 5: What are the Definite Predictions of the
Initial Condition for Features of the Universe on Familiar Scales?}

We may treat this problem briefly because the obvious features of the
universe on familiar scales that are traceable to the initial
condition have been discussed qualitatively in Section II. However,
those qualitative conclusions raise quantitative questions:

What are the coarse-grained variables defining a quasiclassical realm
governed by deterministic laws and how are these variables related to
the principle that determines the initial condition? How refined a
quasiclassical description of the universe is possible before
decoherence is lost or determinism is overwhelmed by quantum noise? How far
in space and time can a quasiclassical description be extended? How do
the phenomenological equations of motion that exhibit the determinism of
the quasiclassical realm follow from the fundamental dynamical law, an initial
condition, and an appropriate coarse-grained description? 
What is the connection of the coarse graining used to define
a quasiclassical realm with that which is necessary to exhibit a second law of
thermodynamics?  How far out of equilibrium is the early universe in
this coarse graining?

In short, a theory of the initial condition presents the challenge of
defining {\it quantitatively} those features of the universe on familiar
scales which are traceable, in part, to the nature of the initial
condition.

\noindent{\bf $\bullet$ Problem 6: What are the Definite Predictions of the 
Initial Condition on Microscopic Scales}

Our understanding of the world on microscopic scales above that set by
the Planck length is summarized by the effective field theories which 
govern phenomena on these scales --- for example, the standard model
of elementary particle physics. However the forms of these effective
field theories 
may be only distantly related to the form of the fundamental dynamical law.
An analogous situation at a different scale may help explain why:
The form of the Navier-Stokes equation which governs the dynamics of
much of the quasiclassical realm is not easily guessed from the 
Lagrangian of the standard model of particle physics. In particular, 
the Navier-Stokes equation incorporates dissipation and depends on
constitutive relations between density, pressure, temperature,
viscosity, etc. --- relations not contained in the Lagrangian of the
standard model.

Of course, we understand qualitatively the relation between the laws of
the standard model and the Navier-Stokes equation. The
Navier-Stokes equation applies, not generally, not
exactly, but only approximately in
particular circumstances.  It is {\it effective} equation with
a limited range of approximate validity. In quantum mechanics particular circumstances
are represented by the quantum state and a coarse-grained description.  It is the quantum state whose
special properties allow a classical approximation,  set up the
conditions for dissipation, determine the constituents, and allow for the
local equilibrium from which the constitutive relations follow.

But the standard model
itself may be only an effective approximation to a more fundamental
dynamical law such as heterotic superstring theory or non-perturbative
quantum gravity.  We may therefore restate the problem of the definite
predictions of the initial condition as follows: {\bf
What features of the effective dynamical laws  that
govern the elementary particle system at accessible energy scales are
traceable to the cosmological initial condition and what to the
fundamental dynamical law?} For instance, what is the origin of the 
locality of the
effective interactions in a theory of the quantum state that is
intrinsically non-local? 

The investigations of the effects of wormholes by Hawking, 
Coleman, Giddings and Strominger, and others 
indicate just how strong the effect of the initial condition on the
effective interactions could be (for a review see, {\it e.g.} \cite{Str91}).  Suppose that the sum over geometries
defining the ``no-boundary'' wave function in (\ref{threetwo})
includes a sum over wormhole
geometries --- four dimensional geometries with many ``handles'' rather
like a teacup has a handle.  Suppose that the Planck scale
(\ref{threethree}) is the characteristic size of
these wormholes in the geometries that contribute the most to the sum.
Fields
propagating in such  geometries can go down a wormhole and emerge from
one. On the much larger scales accessible to us, we would see the effect
of Planck scale wormholes as local interactions which create and destroy
particles. The net effect is to add to any local Lagrangian
an infinite series of local interactions with coupling constants that are  not
fixed once and for all by the fundamental dynamical law or even by a
renormalization procedure, but rather
vary probabilistically with a distribution determined by the
initial condition.  If the distribution was sharp (as was hoped for the
cosmological constant) then the couplings would be predicted. 

A similar decoupling between the observed coupling constants and the
basic Lagrangian would hold if the initial condition predicted domains of space
much larger than our visible universe in which breaking of the symmetries 
of the fundamental dynamical law occurred in different ways in different places
 leading
to a differing effective theories in different domains. The form
of the effective theory governing our domain would then be only
a probabilistic prediction of the fundamental dynamical law.

It has proved difficult to push such ideas very far, 
but their lesson is clear.
The form and couplings of the effective
interactions at accessible scales may be probabilistically distributed
in a way which depends on the initial condition.  Finding these
distributions and how sharp they are is  therefore an important
problem in quantum cosmology.
 
\noindent{\bf $\bullet$ Problem 7: What Does Quantum Cosmology Predict
for IGUSes?}

Most of predictions of the initial condition that we have considered so far
are 
described in terms of alternatives of the
quasiclassical realm.  But there are many sets of decohering histories
of the universe arising from a theory of its initial condition and
dynamics that have nothing to do with the usual quasiclassical realm. 
These sets may be quantum mechanically incompatible with
each other and 
with the usual quasiclassical realm in the sense
that pairs of them cannot be combined into  a common decohering set.  Such
incompatible sets are not contradictory; rather they are complementary
ways of viewing the unfolding of the initial condition into alternative
histories.  The quantum mechanics of closed systems does not distinguish
between such incompatible sets of alternative histories except by
properties such as their classicality. All are in principle available
for the process of prediction.

Yet, as observers, we describe the universe almost exclusively
in terms of the familiar variables of classical physics. What is the
reason for this narrow focus in the face of all the other non-quasiclassical
decohering sets of alternative histories? 
Some see this disparity between the possibilities allowed by
quantum theory and the possibilities utilized by us
as grounds for augmenting quantum mechanics by a further
fundamental principle that would single out one decohering set of histories
from all others \cite{DK96}.  
That is an interesting line of thought, but another is
to seek an explanation 
within the existing quantum mechanics of closed systems.

Human beings, bacteria, and certain computers, are examples of
information gathering and utilizing systems (IGUSes). Roughly, an IGUS
is a subsystem of the universe that makes observations and thus acquires
information, makes predictions on the basis of that information using
some approximation (typically very crude) to the quantum mechanical laws
of nature, and exhibits behavior based on these
predictions.\footnote{IGUSes are complex adaptive systems 
in the context of quantum mechanics.  For more on the general
characterization of complex adaptive systems see \cite{Gel94}.} To
explain why IGUSes are exhibited by the universe, or why they behave the way
they do, or to answer questions like ``Why do we utilize quasiclassical
variables?'', one must seek to understand how IGUSes evolved as physical
systems within the
universe.  In quantum cosmology that means examining the probabilities
of a set of histories that define alternative evolutionary tracks.  For
IGUSes that can be characterized in terms of alternatives
of the usual quasiclassical realm, it is a plausible conjecture that
they evolved to focus on the usual quasiclassical alternatives because
these present enough regularity over time to permit prediction by
relatively simple models (schemata). This would be one kind of
explanation of why we utilize the usual quasiclassical realm. However, we should
not pretend that we are close to being able to carry out a calculation
of the relevant probabilities or even likely to be in the early 21st
century!

But what of sets of histories that are completely unrelated to the usual
quasiclassical realm?  Might some of these exhibit IGUSes with high probability
that make predictions in terms of variables very different from
the familiar quasiclassical ones? Or is the usual quasiclassical realm somehow
distinguished with respect to exhibiting IGUSes? To answer such
questions one would need a general characterization of IGUSes 
that is applicable to all kinds of histories --- not just quasiclassical ones ---
and an ability to calculate the probabilities of various
courses of the IGUSes'  evolution. Such questions, while quite beyond
our power to answer in the present, illustrate the range of predictions
in principle possible in a quantum universe from a fundamental theory
of dynamics and the initial condition of the universe.

\section{Unification}

The universal laws that govern the regularities of every physical
system are one goal of physics.  A fundamental dynamical law is one objective.
Quantum cosmology is concerned with the equally necessary
fundamental law specifying the initial condition of the universe.

Historically, many of the advances towards the fundamental laws have had in
common that some idea that was previously thought to be universal
was subsequently seen to be only a feature of our special place
in the universe and the limited range of our experience.  With more
data, the idea was seen to be a true physical fact, but one which is a
special situation in a yet more general theory. The idea was 
a kind of ``excess baggage'' which had to be jettisoned to reach a
more general, more comprehensive, and more fundamental perspective 
\cite{Har90b}. 

It is not difficult to cite examples of such excess theoretical baggage
in the history of physics:
the idea that the earth was the center of the universe, the idea of
Newtonian
absolute time, the idea that the increase entropy was a basic dynamical
law, the idea that
spacetime geometry is fixed, the idea of a classical world separate
from quantum mechanics, {\it etc.~etc.} Further, and more importantly for the
present discussion, one can cite examples concerning the nature of the
fundamental laws themselves: the idea that thermodynamics 
was separate from mechanics, the idea that electricity was
separate from magnetism, and more recently the idea that
there were separate weak and electromagnetic interactions.
These seemingly distinct theories were eventually unified. 
Today, extrapolations of the standard model of the electro-weak and strong
interactions suggest a unified theory of these forces characterized by
an energy scale a little below the Planck scale.

Examples such as those just cited have led some physicists to speculate
that the existing separation between the dynamical laws for the
gravitational and other  forces is also an  example of excess baggage
arising from the limitations of present experiments to energies well below the
Planck scale, 
and to search for a unified fundamental
law for dynamics of {\it all} the forces.  Secure in the faith that fundamental laws are
mathematically simple, heterotic superstring theory or its extensions
have been the inspiring results. 

However, such a unified {\it dynamical} law does not really deserve the
common designation of  ``a theory of everything'' 
or a ``final theory''. Quantum cosmology offers a further opportunity for unification beyond
dynamical laws.  Could it be that the apparent
division of the fundamental laws into a law for dynamics and a law for
an initial quantum state is a kind of excess baggage similar to those
described above?  Gell-Mann \cite{Gel89} has stressed that there is
already an element of unification in ideas such as the ``no-boundary''
proposal.  In (\ref{threetwo}) the same action that determines
fundamental dynamics also determines the quantum state of the universe.
Despite this connection, 
the ``no-boundary'' proposal is a separate principle specifying  one
wave function out of many possible ones. 
Thus we have an eighth
problem for quantum cosmology:

\noindent {\bf$\bullet$ Problem 8:  Is there a 
Fundamental Principle that would Single Out {\it Both} a
Unified Dynamical Law {\it and} a Unique Initial Quantum State for the Universe?
 Could that same
Principle Single Out the Form of Quantum Mechanics from Among Those Presented
by Generalized Quantum Theory?}

In such a unification of the law of dynamics, the cosmological boundary
condition, and the principles of quantum mechanics, 
we would, at last, have a truly unified fundamental law of
physics governing the universe as a whole and everything within it.
That is truly a worthy problem for physics in the twenty-first century!

\section{Further Reading}

Ref \cite{Halxx} is  {\sl Scientific American} article introducing quantum cosmology.
 An accessible but more advanced introductory review is
\cite{Hal91}. That article contains a nearly exhaustive list of references
at the time and a guide to the literature. For an introduction to the quantum mechanics of closed systems,
see {\it e.g.}\cite{Har93a}. For an exposition of the applications of quantum
mechanics to cosmology see \cite{Har95c}.

\acknowledgments

Thanks are due to G. Horowitz and J. West for critical readings of
the manuscript. 
Preparation of this essay was supported in part by the US National
Science Foundation under grants PHY95-07065 and PHY94-07194.


\begin{references}
\bibitem[]{} 
No attempt has been made to assemble a complete list of references to
the topics touched on in this essay. Rather, the emphasis has been
placed on reviews of topics that are 
most useful directly to the points considered.  These are not always the
earliest nor the latest references.
\end{references}
\end{document}